# Piloting Virtual Reality Photo-Based Tours among Students of a Filipino Language Class: A Case of Emergency Remote Teaching in Japan

**Roberto Bacani Figueroa Jr***
University of the Philippines Open University, Philippines
*corresponding author: *rbfigueroa1@up.edu.ph*

**Florinda Amparo Palma Gil**
Tokyo University of Foreign Studies, Japan

**Hiroshi Taniguchi**
University of the Philippines Open University, Philippines

Received 6 September 2020; accepted 28 January 2021; published 24 August 2022.

**Abstract**

The State of Emergency declaration in Japan due to the COVID-19 pandemic affected many aspects of society in the country, much like the rest of the world. One sector that felt its disruptive impact was education. As educational institutions raced to implement emergency remote teaching (ERT) to continue providing the learning needs of students, some have opened to innovative interventions. This paper describes a case of ERT where Filipino vocabulary was taught to a class of Japanese students taking Philippine Studies in a Japanese university using a cognitive innovation based on virtual reality, an immersive technology often researched for immersion and presence. Students were divided into three groups to experience six lessons designed around virtual reality photo-based tours at different immersion levels. While the effect of immersion on satisfaction was not found to be statistically significant, presence and satisfaction were found to be correlated. Despite challenges that were encountered, benefits like enjoyment, increased engagement, and perceived learning were reported by the students. Our findings exemplify how emerging multisensory technologies can be used to enhance affective and cognitive dimensions of human experience while responding to gaps created by the spatial limitations of remote learning.

## 1. Introduction

The COVID-19 pandemic has greatly changed societies in almost every aspect of life since the end of 2019. One area that has been heavily affected by this global crisis is education. The year 2020 served as a turning point for educators across the globe, causing them to reevaluate their pedagogy and look for innovative ways to overcome the challenge of limiting in-person meetings for teaching and learning. The effects varied depending on the subject being taught, the cultural setting, and the technologies available. This paper hopes to contribute to this fast-growing body of knowledge on context-specific innovations by presenting a study that describes the process of prototyping a technology-based innovation in the context of what is known in the educational literature as emergency remote teaching (ERT), which was defined as the sudden and temporary shift of delivery to a distance or remote mode of instruction as a result of extremely disruptive phenomena (Hodges et al., 2020; Mohmmed et al., 2020). Results from this study may be of interest not only to researchers in the field of education, but also in other fields like psychology, information technology, and cognitive science as they have implications on the dynamics of immersion, presence, and satisfaction in multi-sensory experiences brought about by immersive technologies.

### *1.1. Foreign Language Teaching in Japan*

Popular foreign language learning methods in Japan include in-person lessons at private language schools, online one-on-one lessons, and the use of language applications on mobile phones or personal computers. However, what is referred to as foreign language learning, mostly refers to learning the English language. While the English language is taught from the elementary level to the university level, with grammar and translation methods dominating over others, the teaching of foreign languages other than English is mostly unheard of (Butler, 2007). Among languages that were rarely taught in Japan is Filipino, the national language of the Philippines.

Among Japanese universities, Osaka University and Tokyo University of Foreign Studies are the only ones that offer a full major course in the Filipino language as of this writing. With this, comes the scarcity of teaching and learning materials that are content-based (Laranjo, 2020), especially those targeting Japanese learners. These universities have facilitated Filipino language education using traditional in-person classroom-based or blended pedagogy using a learning management system (LMS). Learning activities that would provide a more immersive experience for learners and establish relevance and context of the target language were offered in the form of study abroad programs or study tours embedded within the four-year curriculum.

### *1.2. The Situation of Education in Japan at the Dawn of the Pandemic*

Just like the rest of the world, university teachers and students in Japan did not expect the effect that the pandemic would bring to the academic year 2020. When the COVID-19 outbreak occurred, the Japanese government called a nationwide state of emergency and restricted people's mobility entering and leaving Japan. As a result, foreign students who returned to their home countries during the spring break in 2020 were included in the reentry ban (Osumi, 2020). At the same time, 80 to 90% of Japanese universities recalled their students studying abroad, giving no guarantee of returning to their programs (Wortley, 2021). According to the result of a nationwide survey released by Japan's Ministry of Education, Culture, Sports, Science, and Technology (MEXT) in May of 2020, as a countermeasure to the COVID-19 outbreak, 930 universities out of 1046 that responded (86.9%), delayed the start of classes and more than 96.6% of the said universities either decided or were considering conducting distance learning using different forms of media. Since the most popular form of education in Japanese universities before the pandemic was face-to-face classes, teachers were forced to hurriedly prepare for switching to online classes during the short delay (Inoue, 2020). On the other hand, incoming first-year students, expecting to start a new active life in their new school, had to attend online classes without being able to set foot on campus and before experiencing face-to-face lectures with their professors (Hirabayashi, 2020).

Although Japanese universities had already been introduced to the use of LMS for language education even before the pandemic, actual usage by teachers, staff, and students was still minimal (Murakami, 2016). Teachers had to make significant adjustments in shifting to online classes. The Mainichi Shinbun, one of the major newspapers in Japan, surveyed 66 institutions and showed one teacher describing how his preparation time has gone up fivefold while some teachers thought of doing online classes as becoming Youtubers or that they should be like radio broadcasters (Mainichi Japan, 2020). As for the students, it was reported that "…while 59.2 percent were positive about online classes, 21 percent said they did not wish to take part, reflecting concerns about the quality of education that remote learning provides and finding the right environment for participating" (Kyodo News, 2020, para 2). In fact, in a survey from an organization petitioning the return of classes to face-to-face, it was reported that *the quality of lessons was lower than face-to-face* was the second highest answer given by 845 out of 1,500 respondents, regarding the perceived disadvantages of an online class (DaigakuIkitai, 2020). Even at the University of Tokyo, which was the fastest to adapt to online classes, 70% of the students who were surveyed said that online classes were mere approximations of regular classes (Shoji, 2020). However, an exploratory study on first-year students' perception of university online lessons reported that the efficiency of technological innovation brought about a positive response to online learning as well as changed t*he feeling of loneliness of in-home study* to *being able to concentrate on individual learning efficiently through online lessons* (Hirabayashi, 2020).

Teachers' lack of knowledge in developing efficient online classes seemingly presents a challenge in achieving the fourth sustainable development goal, which is about ensuring equitable and quality education for everyone. However, it also presents an opportunity to explore, exploit, and explain innovative ideas for revolutionizing education that would address the gaps and restrictions presented by the current situation (Kitakka, 2020).

### *1.3. Virtual Reality Technology in Emergency Remote Teaching (ERT)*

Virtual reality (VR) is a system of specific hardware and software that brings about a real-time, computer-generated three-dimensional environment that users, having a perceived self-location, can navigate and interact with (Hayward, 1993). Despite having many features, two were found to be common among those investigated by VR researchers: *Presence* and *Immersion*. Presence or telepresence is the *feeling of being there* in a technology-mediated environment (Heeter, 1992), while immersion is defined as the aspects of hardware and software systems, such as display quality and stereoscopy, which could facilitate presence (Slater & Wilbur, 1977).

VR has gained increasing popularity in technology-integrated classrooms as reported in studies done by Chang et al. (2016), Coyne et al. (2018), and Kim et al. (2001). The ability of VR to immerse learners in a virtual environment or simulate a geographical location without having to go there is potentially useful in addressing the gap caused by limited mobility among learners due to the lockdowns caused by the pandemic around the world. One instance of a VR application that provides opportunities to teachers without a programming background is the virtual reality photo-based tour or simply VR tour. VR tours are interactive tours based on 360 photos of real-world locations with hot spots around the environment that provide information about an area or objects. Equipment and devices for 360 photos have become more affordable in recent years.

Furthermore, affordable and easy-to-use platforms for creating interactive tours using these media have become widely available. The lockdowns that caused the lack of access to location-based educational activities have increased the potential use of VR tours in education and other learning contexts, especially among courses and subjects that would have involved field trips in normal cases. However, reports were scarce during the time that this study was conceptualized.

Even rarer were studies on remote teaching and learning methodologies for the Filipino language. The few papers found on teaching Filipino as a foreign language mainly focused on the details of the curriculum and the background of students. Only some of these papers mentioned the methodology of teaching and the challenges of teaching Filipino - Quirolgico-Pottier (1997), Luquin (2016), Pambid-Domingo (2010), Barrios-Le Blanc (2010), and Laranjo (2020). Pambid-Domingo (2010) mentioned the use of authentic materials, videos, and songs, while Barrios-Le Blanc (2010) explained teachers' techniques in using poetry for teaching. However, none of them investigated using information and communication technology (ICT) in teaching the subject matter among Japanese students. This could be brought about by the general apprehension among teachers in Japan about using ICT-based novel interventions (Teaching and Learning International Survey [TALIS], 2018).

According to the 2018 Teaching and Learning International Survey (TALIS) survey by the Organization for Economic Co-operation and Development, Japan ranked 45 out of 50 on teachers' preparedness in using ICT for teaching. Fewer teachers felt "well pre-pared" or "very well prepared" in ICT for teaching in Japan compared to other participating countries. While Japanese universities have learning management systems, they are used for less than 20% of all courses. Moreover, due to a lack of staff for creating digital content and maintaining ICT systems, insufficient ICT skills for both academic and general staff, and a limited budget, ICT tools were mostly used for syllabus systems (89%), student information systems (63%), and campus WiFi networks (79%), impeding the introduction and promotion of ICT tools for educational use (Funamori, 2017).

A few months into the pandemic, news articles have shared how ICT was used by various learning institutions in Japan to fill the gap created by the state of emergency caused by the COVID-19 outbreak (O'Donoghue, 2020). However, the authors have not yet found any study that outlined how a particular ICT-based intervention was designed and how it affected the students' learning experience and outcome. This is also true even in foreign language teaching and learning, much more so among less popular languages like Filipino. Aside from a recent study reporting some motivational effects of VR tours in a university in Japan (Figueroa et al., 2020), no article regarding VR tours for teaching the Filipino language in Japan in the context of remote teaching has been found as of this writing. Thus, when the state of emergency was first announced in the country, and a shift to online classes was decided by the university administration, an exploratory study on utilizing VR tours in teaching vocabulary to zero-beginner learners of the Filipino language was carried out.

### *1.4. Aims of The Study*

This study aimed to provide insights as to how an intervention using VR tours could have an impact on Japanese students of a Filipino Language class in the context of ERT during the time of the COVID-19 outbreak in Japan. More specifically, it aims to answer three research questions (RQ):

> RQ1: How different are the levels of satisfaction between students who experienced VR tours with varying levels of immersion?
>
> RQ2: How are the levels of satisfaction among students related to presence in the VR tours?
>
> RQ3: What are the other benefits of incorporating VR tours in an online Basic Filipino Language Class designed for Japanese students?

## 2. Materials and Methodology

This study investigated the impact of VR tours on Japanese students of a Filipino Language class in the context of ERT during the time of the COVID-19 outbreak in Japan using both quantitative and qualitative analysis. In the quantitative analysis, descriptive statistics and non-parametric tests were used, while in the qualitative analysis, reflexive thematic analysis was used.

### *2.1. Virtual Reality Photo-based Tours (VR Tours)*

The researchers adopted six VR tours into a Filipino Language Course for beginners at a Tokyo-based university over six lessons from May 22 to June 26 of 2020. Each VR tour was inserted into a 90-minute online synchronous lesson. The aims of each tour were fourfold: (1) to increase students' vocabulary, (2) to develop students' ears for recognizing Filipino sounds and rhythm, (3) to let them experience the Philippines albeit virtually, and (4) to let them experience a new way of learning a language. By the end of the six VR tours, the students were expected to be able to introduce popular Japanese tourist spots in a VR tour of their own using the vocabulary that they had learned from the tours and to be able to introduce the places using the correct pronunciation and rhythm.

Each tour introduced a place in the Philippines while featuring seven new Filipino words. The tours were given titles based on their main location of interest, and the seven words introduced in each tour were related to the things, people, places, and events that could be found in these locations. Figure 1 shows the tour used for Lesson 6, which was entitled *Plasa*. A hyperlink to the interactive online tour is also provided as a note for those interested in experiencing it.

**Figure 1**

*Lesson 6: Plasa Virtual Tour*

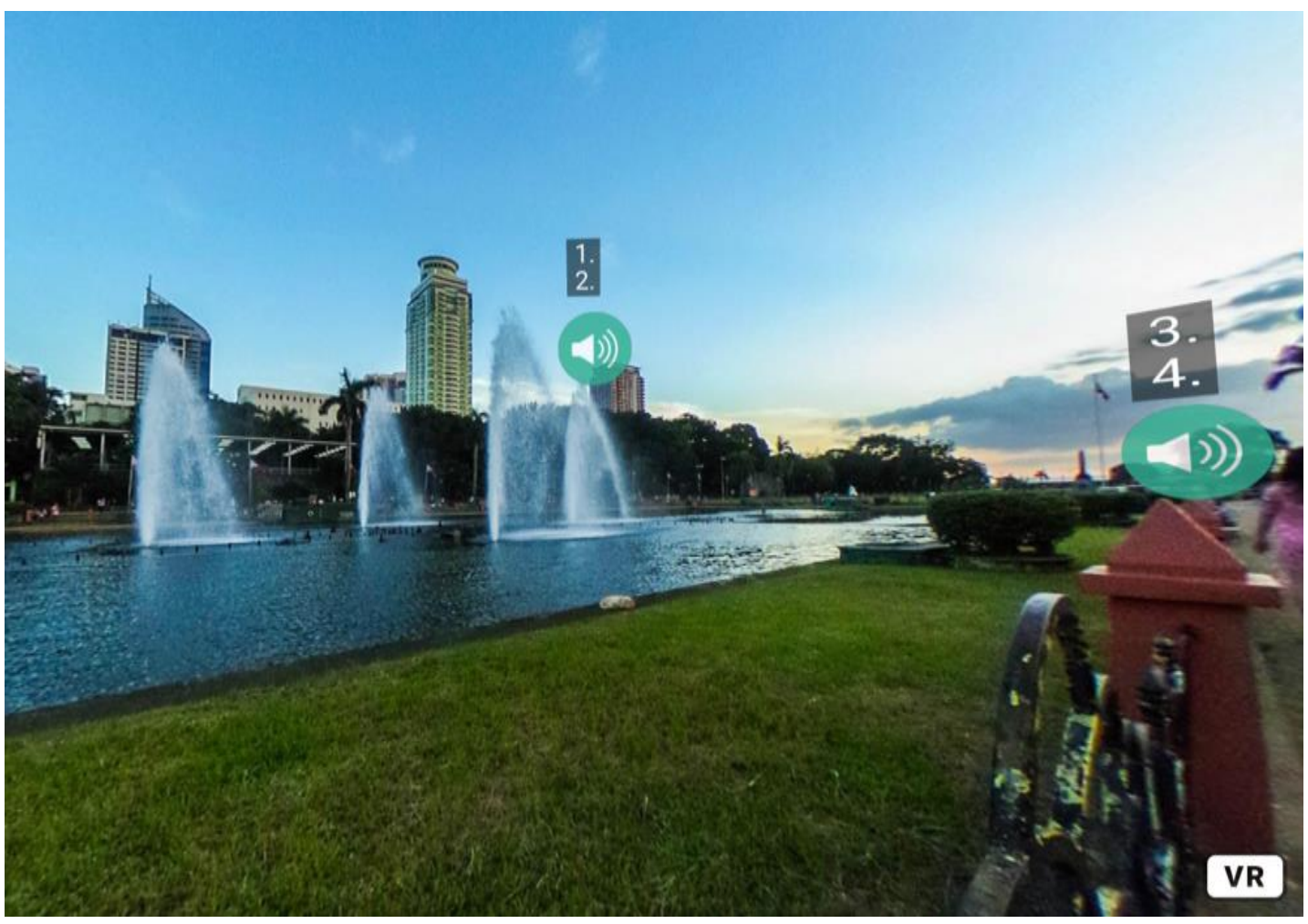

*Note.* URL: *(https://figtreeacademy.org/vr/plaza/)*

Table 1 shows a detailed list of the tours for reference. The first, second, fourth, and sixth tours had audio guidance in Japanese and Filipino by two narrators (a Japanese native speaker and a Filipino native speaker). In contrast, the third and fifth tours were narrated purely in Filipino by a Filipino native speaker.

**Table 1**
*The VR Tours*

| Number | Title | Date | Language |
|---|---|---|---|
| 1 | Rizal Park (name of a park) / Pilot Tour | May 22 | Japanese and Filipino |
| 2 | Aurora (name of a province) | May 29 | Japanese and Filipino |
| 3 | Quiapo Church (name of church) | June 5 | Filipino |
| 4 | LRT (light railway transportation) | June 12 | Japanese and Filipino |
| 5 | Museo (museum) | June 19 | Filipino |
| 6 | Plasa (plaza) | June 26 | Japanese and Filipino |

### *2.2. Participants*

All six lessons were conducted among a class of 15 first-year Japanese students majoring in Philippine Studies. All students in the class volunteered to participate in the study.

### *2.3. Data Collection*

Data collected were the following: pre-VR Tour survey, post-semester survey, six after-class surveys, two focus group discussions, and observation notes. Survey questions can be found in Appendix A1 and A2; focus group discussion questions are in Appendix B; and observation notes are in Appendix C.

The pre-VR tour survey was used to decide the groupings. Random selection was carried out among learners who had compatible phones. The participants were divided into three groups - *high immersion group, moderate immersion group,* and *low immersion group*. Quantitative items from five out of six after-class surveys were used for answering RQ1 and RQ2. Open-ended questions from six after-class surveys were used for answering RQ3 and the results were compared to the data collected from the post-semester survey, two focus group discussions, and observation notes.

### *2.4. Procedure*

As shown in the procedural diagram in Figure 2 below, the study was implemented in several phases. As part of the preparation phase, students were asked to accomplish a *pre-VR tour survey***.** Students were assigned to three groups based on their phone's specifications revealed by their answers to the *pre-VR tour survey*.

**Figure 2**

*Procedural Diagram of the Study*

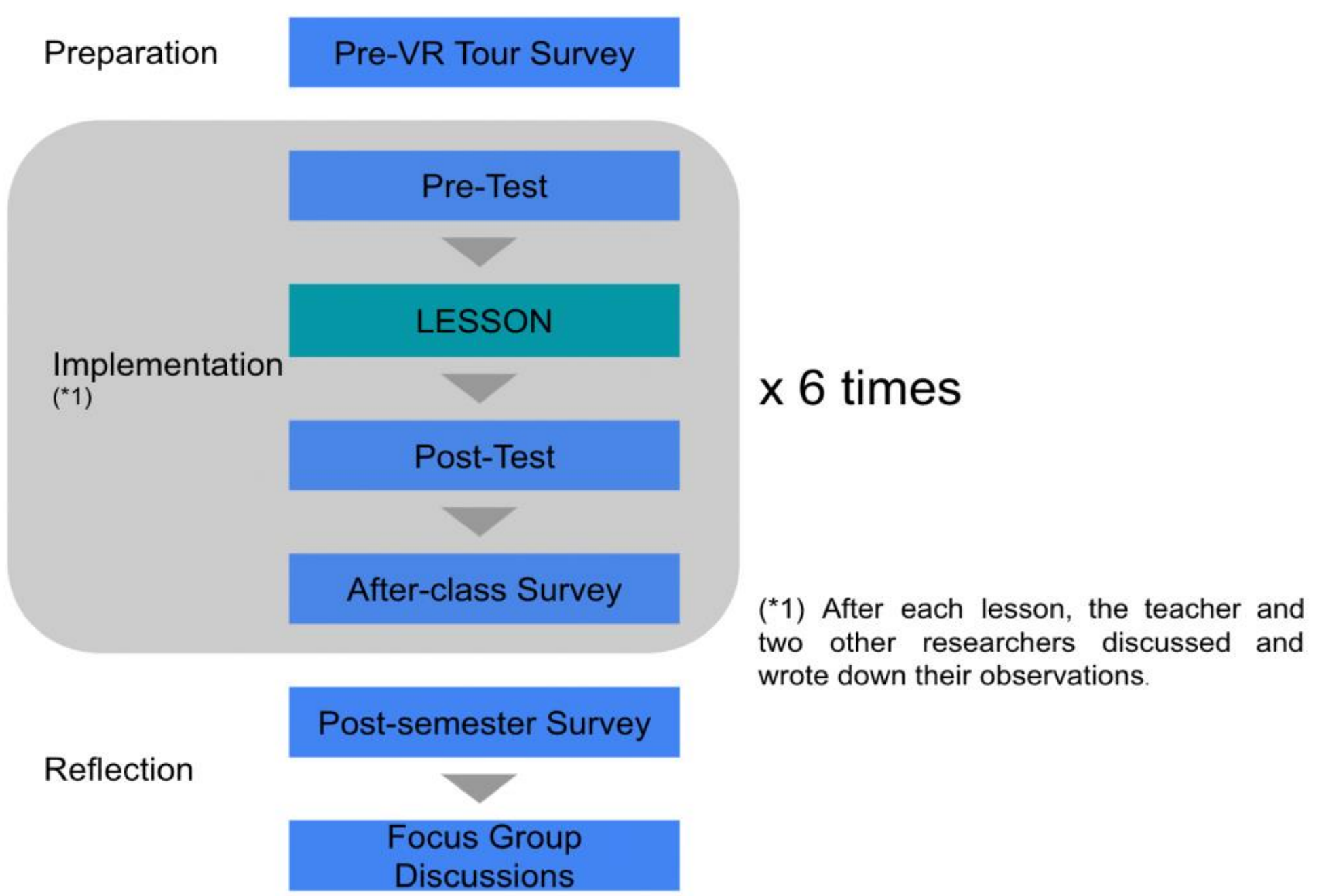

Six lessons were conducted with groups following the same lesson structure. At the beginning of the lesson, students were asked to do a *pre-test of target vocabulary words*. Each virtual tour served as one of the main activities of each lesson. Lessons were structured into four phases [Lesson Introduction → Virtual Tour → Focus on Grammar → Application].

In the *Lesson Introduction* phase, the theme and the objectives of the lesson were introduced. In the *Virtual Tour* phase, a new set of vocabulary related to the lesson's theme was introduced using a VR tour. Students assigned to the *high immersion group* participated in all the VR tours using their phones and VR goggles delivered to their homes. Those assigned to the *moderate immersion group* participated in all the VR tours using their computers or smartphones, but without using VR goggles. Those assigned to the *low immersion group* viewed and listened to PowerPoint-based tours that had the same content as the VR tours but used 2D photos. During this phase, the teacher reminded the students to experience the VR tours three times for vocabulary reinforcement.

This was followed by the *Focus on Grammar* phase, where new grammar was taught. In the *Application* phase, students were asked to use the presented vocabulary words and new grammar to form sentences describing sample photos provided by the teacher. Finally, a post-test containing items related to the presented vocabulary words was given, followed by an *after-class survey*.

After conducting lessons 1, 2, and 3 (Rizal Park, Aurora, and Quiapo Church), the students were divided into pairs. They were asked to make their own tour introducing a place in Japan by using 2D photos and Google Slides. They were allowed to choose the language they would use as long as they included seven Filipino words from the three tours. Likewise, after lessons 4, 5, and 6 (LRT, Museo, and Plasa), the students were again divided into pairs. They were asked to make their own VR tour using the platform *Story Spheres,* and VR photos of places in Tokyo which were provided to them. They were given instruction to use seven words from the last three tours and the vocabulary words learned within the semester.

Another *survey* was conducted at the end of the semester, and from the results, six students were invited for two *focus group discussions.* Two students from each group were chosen to join the FGDs. The students who were interviewed were the ones who mentioned something related to the VR tours in their answers in the post-semester survey. After each lesson, the teacher and two other researchers discussed and wrote down their *observation notes*.

### *2.5. Iterative Improvement based on Challenges Encountered*

Strategic integration of VR tours in class involved iterations of creative thinking, brainstorming, testing, evaluation, revision, and synthesis -activities that characterize a cognitive innovation. Like in most innovations, several challenges were encountered during implementation. These challenges were addressed by adjusting the design and implementation of succeeding lessons described briefly below. However, other challenges were not addressed due to time constraints and contextual limitations.

**Device Compatibility.** The students were grouped according to the compatibility of their mobile devices with VR Tours and their willingness to try VR Tours. However, two participants in the high immersion group had incompatible phones. Hence, the groupings had to be adjusted after the first lesson resulting in the decision to collect quantitative data from lessons 2 to 6 only.

**Device and Tour Usability.** An orientation was conducted since it was the first time for many of the participants to use VR tours. However, high immersion group participants continued to have difficulty using the device until the second tour. They started becoming comfortable only from the third tour with careful guidance and reminders in operating the tours.

**Tour Technical Problems.** Participants from all three groups experienced technical difficulty in loading the audio of the tours, most especially the high immersion group participants. Hotspots were supposed to activate certain audio explanations when gazed on or clicked. Some of the audio files were slightly larger, which caused them to be loaded rather slowly, especially when the student's internet connection was intermittent. This interrupted the flow of the lessons significantly. However, these were resolved by restarting the application or refreshing the page, or by using a different mobile device.

**Tour Language.** Since the participants were Japanese speakers who generally had no exposure to the Filipino language, it was decided to have the first tour be explained in Japanese with highlights on Filipino terms that were the target words for that lesson. Labels corresponding to the target words were also placed on the hotspots. However, this removed any challenge in the task and lessened the impact of the tour in terms of student engagement. Thus, the succeeding tours did not have textual labels, which turned out to be useful in training the students' listening skills. On the other hand, the two tours that had audio in all Filipino increased the difficulty of the task immensely. As a response, the teacher translated and explained the script's meaning after the third listening part so students could consult the script as often as they wanted to make sense of what the tour guide was saying in each hotspot.

## 3. Data Analysis

Quantitative and qualitative analyses were performed to answer the research questions of the study. In this section, we discuss the three research questions and the analyses performed to answer them. All quantitative procedures were carried out using *R*, an open-source software offering libraries for statistical analysis (R Core Team, 2012). R-Studio, a commonly used integrated development environment for R, was utilized for easy documentation and organization of data.

### RQ 1: Levels of Immersion and Satisfaction Ratings

Participants' satisfaction ratings collected from *after-class surveys* of Lessons 2 to 6 were analyzed and grouped by immersion level. The omnibus test used to determine the effect of immersion level on satisfaction was the Kruskal-Wallis test as a non-parametric alternative to the analysis of variance for the following reasons: 1) There was a small number of participants who were randomly allocated to the treatment groups (i.e., *moderate* and *high immersion groups*); 2) Samples were mutually independent; and 3) Dependent variables were at least in the ordinal scale. Boxplot visualizations were generated using the *ggplot2* library (Wickham, 2016), while the Kruskal-Wallis function was performed using the stats library (R Core Team, 2012).

### RQ 2: Presence and Satisfaction Ratings

This section describes the tests performed and plots that were visually inspected to answer research question 2. Participants' satisfaction and presence ratings collected from *after-class surveys* of Lessons 2 to 6 were analyzed. Scatterplots of satisfaction and presence ratings were generated in five lessons to visualize the monotonic relationship between the two variables using the *ggplot2* library (Wickham, 2016). Kendall's tau was calculated to determine the correlation between satisfaction and presence ratings based on the assumptions: 1) Data from the paired observations appeared to follow a monotonic relationship, and 2) variables were measured at least in the ordinal scale. Furthermore, Kendall's tau was chosen over Spearman's rho because the p-values of the former were found to be more accurate with smaller sample sizes (Marshall & Boggis, 2016), which was the case in the current study.

### RQ3: Benefits of VR tours

The qualitative analysis employed in this study was reflexive thematic analysis. To execute the analysis, data were collected from the six *after-class surveys*, the transcription of the two *FGDs*, and the *observation notes.* These data were collected to answer RQ3: *What are the other benefits of incorporating VR tours in an online Basic Filipino Language Class designed for Japanese students?*

The data analyzed from the six *after-class surveys* were the reasons given by the participants regarding the rating of their experience of trying each VR Tour. The reasons were categorized into codes and sub-themes until two main themes emerged from the categorization. After this process, the codes and sub-themes were triangulated with the answers of the six participants invited to the FGDs, results from the post-semester survey, and observation notes of the authors.

Categorizing the reasons given by the participants regarding the rating of their VR Tour experience produced 20 codes which were further categorized into 7 sub-themes and finally into two main themes – the **Benefits of VR Tours** and the **Problems and Challenges encountered** while doing the VR Tours.

## 4. Results

Notable findings from analyses that were carried out are organized according to the three research questions.

### RQ 1: Levels of Immersion and Satisfaction Ratings

The boxplots in Figure 3 visually illustrate the median and interquartile range of satisfaction ratings given by the three groups. Outliers were found in the *high immersion group* in the second lesson and the sixth lesson, and the *moderate immersion group in* the fifth lesson.

**Figure 3**

*Satisfaction Boxplots of Three Immersion Groups per Lesson*

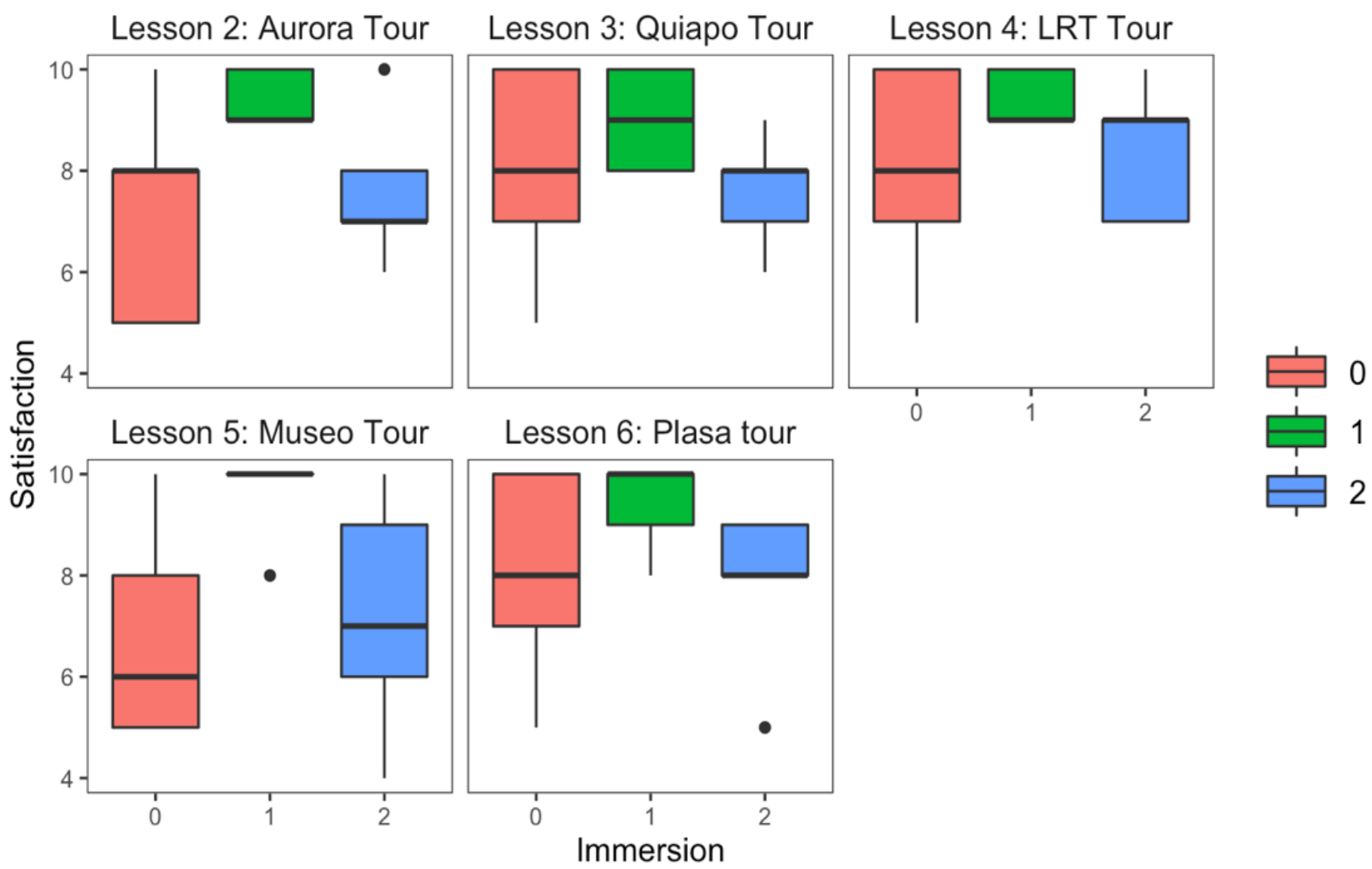

**Figure 4**

*Scatterplots of Satisfaction and Presence Ratings in Five Lessons*

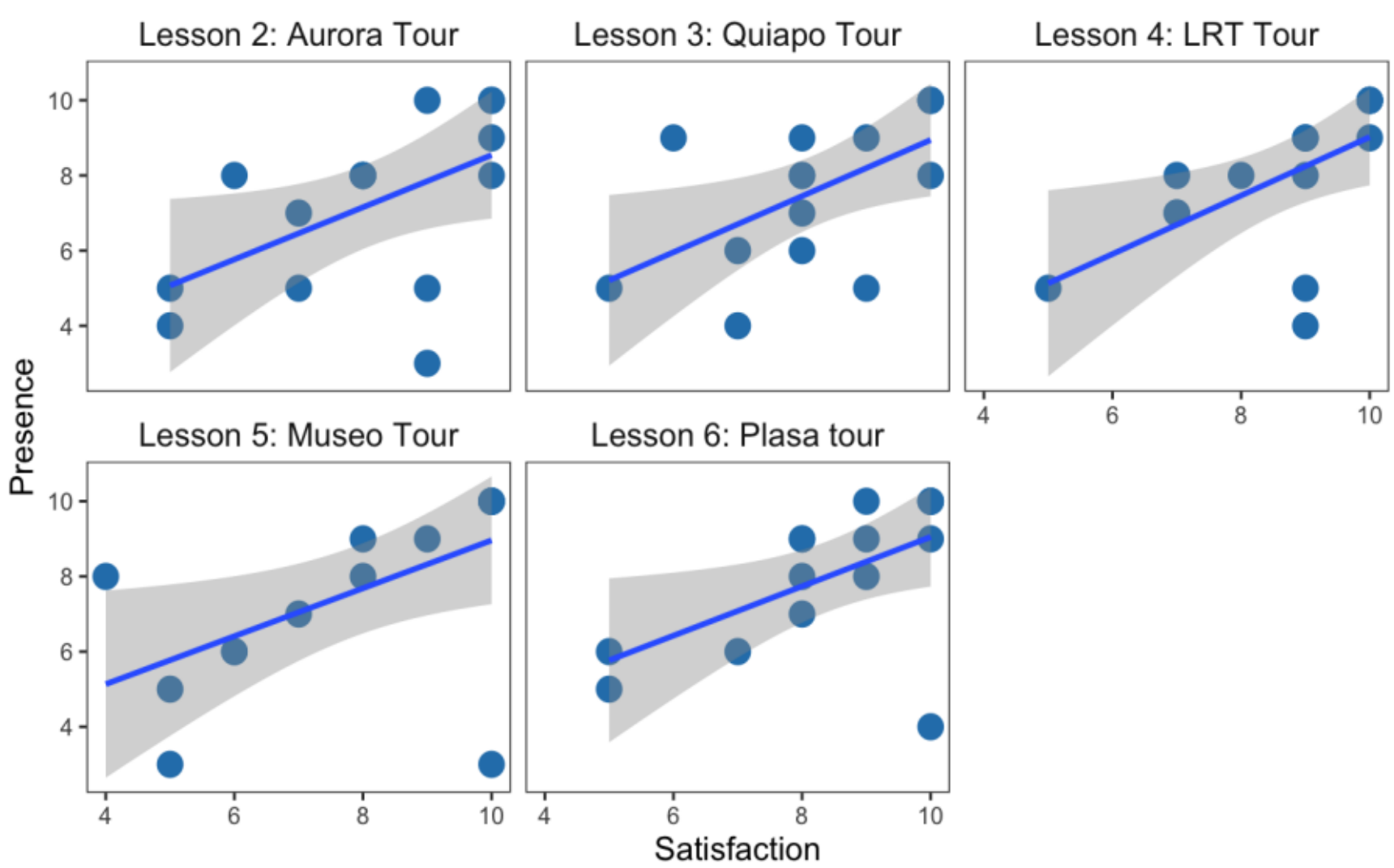

The medians of satisfaction ratings by the *low immersion, moderate immersion,* and *high immersion groups* in the second lesson were 8 (IQR = 5 - 8), 9 (IQR = 9 - 10), and 7 (IQR = 7 - 8), respectively. The medians of satisfaction ratings by the *low immersion, moderate* *immersion,* and *high immersion groups* in the third lesson were 8 (IQR = 7 - 10), 9 (IQR = 8 - 10), and 8 (IQR = 7 - 8), respectively. The medians of satisfaction ratings by the *low immersion, moderate immersion,* and *high immersion groups* in the fourth lesson were 8 (IQR = 7 - 10), 9 (IQR = 9 - 10), and 9 (IQR = 7 - 9), respectively. The medians of satisfaction ratings by the *low immersion, moderate immersion,* and *high immersion groups* in the fifth lesson were 6 (IQR = 5 - 8), 10 (IQR = 0), and 7 (IQR = 6 - 9), respectively. Finally, the medians of satisfaction ratings by the *low immersion, moderate immersion,* and *high immersion groups* in the sixth lesson were 8 (IQR = 7 - 10), 10 (IQR = 9 - 10), and 8 (IQR = 8 – 9), respectively. The medians and interquartile ranges presented in the boxplots also revealed that students in the *moderate immersion group* generally gave higher satisfaction ratings compared to the *low* and *high immersion groups.*

However, no statistically significant differences were found among three groups who experienced tours in the second, [H(2) = 4.52, p = .11]; third, [H(2)= 2.49, p = .39]; fourth, [H(2)= 1.63, p = .44]; fifth, [H(2)= 4.99, p = .08]; and sixth, [H(2)= 3.08, p = .22], lessons.

**RQ 2: Presence and Satisfaction Ratings**

The scatterplots shown in Figure 4 illustrate the monotonic relationship between the two variables in five lessons. Results revealed that there was a statistically significant strong correlation (N=15) between presence and satisfaction ratings in the second ($\tau b$ = .50, p=.02), third ($\tau b$ = .51, p=.02), fourth ($\tau b$ = .67, p=.002), fifth ($\tau b$ = .65, p=.002), and sixth ($\tau b$ = .56, p=.01) lessons.

**RQ3: Benefits of VR tours**

The codes and categories collected in the thematic analysis are presented in Table 2. Two main themes emerged: Benefits and Challenges. Since the challenges were already discussed in the methodology section as part of the iterative improvement process, this section will discuss the subthemes related to benefits.

**Increased Engagement.** Out of the 20 codes identified, most of the reasons given by the participants for their VR Tour experience ratings alluded to the excitement of doing something new or the feeling of being transported to a different place which made them say that learning was enjoyable. These trends were also the first things that the teacher observed among students who used the VR tours for the first time. Participants of the *high immersion group* and *moderate immersion group* were observed to have eagerly and diligently engaged in learning the new sets of vocabulary words and in fini-shing the activities of each tour. The novelty of the intervention naturally stirred students' interest in participating in each class. Although a few chose to use the conventional PowerPoint presentation-based tours because they said that they were not good with technology or that their mobile phones and PC were not compatible, five out of the six participants of the focus group discussions said that the VR tours were their most en-joyable activity in class. The tours made them feel that they were in the Philippines. One of the participants said, "*I could feel the Philippines with the VR tour even if I'm not there.*" Three out of the five participants even said that the application activity of the course, which allowed them to make their own tours, was extremely fun and made them think about the difficulties encountered by the teacher and researchers in setting up the tours.

**Table 2**

*Codes and Themes Created from the Categorization Process*

| Main Themes | Sub-themes | Codes |
|---|---|---|
| Benefits | Increased Engagement | -Excitement towards a new experience (12)<br>-Excitement in feeling as if in the actual place (10)<br>-Enjoyment in learning (7) |
| | Perceived Vocabulary Retention | -Ease of memorizing words (9)<br>-Ease of understanding (6)<br>-Acquisition of actual usage of new words (6)<br>-Acquisition of new word s (5)<br>-Acquisition of correct pronunciation and rhythm (3) |
| | Interest | -Admiration of the view (16)<br>-Interest in going and seeing the Philippines (3) |
| | Authentic Information about the Philippines | -Acquisition of information about actual places and the actual environment/atmosphere in the Philippines (8)<br>-Acquisition of information about the daily life in the Philippines like rush hour, crowded train, going to church, fashion of people, etc. (8) |
| Challenges | Computer /Phone Trouble | -Trouble in listening to the audio (5)<br>-Trouble in playing the tour smoothly and continuously (4)<br>-Trouble in manipulating or operating the VR Tour (2) |
| | Obscureness of Photo | -Difficulty in understanding the photos (2)<br>-Monotonousness of photos (2)<br>-Unrelatedness of photos to the theme of the tour (2) |
| | Difficulty with the language used for the script | -Disappointment with listening to only in Filipino without Japanese (for some tours) (3)<br>-Disappointment in listening to more Japanese than Filipino (for some tours) (2) |

**Perceived Vocabulary Retention.** The second prevailing theme in the observations and survey results was that students had perceived ease of remembering the words because of the context provided by the virtual tours in the lessons. They recalled that they quickly remembered and understood how the words would be used in the Philippines as they referenced objects in the tour, which made it easy to associate them with their meanings. Many of them associated their ease of remembering the words with the feeling of being in the actual place. This was further validated by the participants of the focus group discussions. A participant commented, "*We can see the real Philippines, and with the actual* *situation, I can learn the vocabulary*." Aside from comments about the benefit of VR Tours for language learning, this study also found comments about the effi-ciency of the program design itself for language learning like, "*the lesson was repeated learning. We learned the vocabulary before VR and during VR. We encountered those vocabulary many times,*" and "*I was able to listen to Filipino sentences. I was able to practice listening to Filipino words and enjoyed learning words.*" From comments like these, the authors learned that the study was able to achieve the aims of each tour.

**Interest in the Subject Matter.** The third prevailing theme was that the VR Tours stirred their interest not just in the activity, but also in the subject matter: the Philippines and its national language, one of the goals of the whole program. As further explained in the latter section, according to the post-semester survey, students with an initially slightly negative view of the country changed their impression of it. The VR tours made them want to visit it more. This was also supported by the frequency of the students' answers regarding the aesthetic value of the tours, which led them to be interested in the lessons as reported in the after-class surveys.

**Authentic Information about the Philippines.** The fourth prevailing theme found from after-class surveys was the authenticity of the information about the Philippines that the participants were able to get from the VR Tours. Participants wrote that they could "*feel the atmosphere"* of the places introduced in the tours, like the church or the train. Half of the participants also mentioned that they were happy to be able to "*learn about the history of the Philippines*", or "*see everyday life in the Philippines*." The authenticity of the information they got from the tours made them feel as if they were walking quietly inside a church full of people praying or as if they were also stuck inside a crowded train. Two focus group discussion participants agreed that "*All tours were fun.*" It was followed by the comment, "*the image about the Philippines made me think I want to go there."*

**Changes in perception of the Philippines.** Aside from the after-class surveys, data were gathered from the post-semester survey. The answers showed important points about the benefits of using VR Tours in a Filipino language class. Answers revealed that some of the students who experienced the VR tours considered the activity one of the most useful activities during the period and that experiencing the tours changed their perception of the Philippines. The six students who wrote about this also said during the focus group discussions that their perception of the Philippines significantly changed. Some of them said that before the tours, they were anxious about visiting the country because they had only associated the Philippines with hot weather and poverty. However, after experiencing the VR tours, they discovered something new. Answers from students like "*I thought many people are poor, but once I saw Manila so developed, I wanted to go*" and "*I am getting interested in the Philippines more and more. I want to visit it*" set the tone of the focus group discussions. They also mentioned that experiencing local and rural life was one of the things that made the tours interesting because what the media had shown them was only about tourist attractions, politics, or poverty. This made them feel more excited to visit the country and learn more about the language to prepare themselves for the actual visit.

## 5. Discussion

The study described in this paper primarily aimed to investigate how integrating VR tours supported a class in Japan that shifted to synchronous online classes as an emergency remote teaching strategy due to the COVID-19 pandemic. The findings of the study unearthed several themes that are worth discussing.

### *5.1. Emergency Remote Teaching as an enabler for Cognitive Innovation*

As Gummerum and Denham (2014) explained, a cognitive innovation starts with an exploratory phase of creatively looking for ideas and probing boundaries, followed by choosing ideas that could be exploited, tested, and improved. It is then followed by reflecting on the experience and synthesizing what has transpired to explain phenomena and generate new questions and ideas. At first glance, one might think that cognitive innovation can only thrive in a more relaxed, unconstrained, and open environment.

However, in this instance, the cognitive innovation of incorporating novel VR tours into an online lesson that involved iteratively brainstorming lesson design and structure-related ideas, selecting which of these ideas to exploit, and constantly reflecting on learning points and things to improve each week, was enabled by limitations brought about by the state of emergency in Japan during the COVID-19 outbreak.

While the term emergency was used in both *emergency remote teaching* and *state of emergency*, its contextual meaning is not comparable to a dying patient being rushed to a hospital or a burning house. Rather, it pertains to a state where certain interventions that were not normally allowed would be considered because of an extremely disruptive phenomenon. For example, in a normal situation, a teacher in Japan would not be allowed to teach a class online in a traditional university. However, because of the state of emergency pronouncement by the government where universities must disallow in-person activities to prevent the spread of COVID-19, conducting classes online through various platforms has become permissible to provide a way for educational institutions to deliver services to their learners. In the same way, because of the sudden closure of borders that prevented learners from learning Filipino vocabulary in a more immersive environment through field trips, VR was considered to be a possible means of allowing learners to somehow experience certain places in the Philippines while learning the language without having to go there.

Before the phenomenon, the idea of using virtual tours in a university setting was not something that could be easily considered as students were able to receive ample stimuli and realistic input from study abroad programs and in-person interactions with native speakers in the physical classroom. The void created by the phenomenon made teachers, learners, and school administrators open to new ideas that would not have been ideal in normal cases, as revealed by reflection conversations held among the observers and the teacher. In this way, temporal and spatial restrictions in education gave way to the openness to new ways of experiencing places while learning and actively participating in ma-king the succeeding iterations better. It also made researchers more attentive to feedback and less afraid to exploit alternatives. The phenomenon was so unique that no previous guideline could have restricted them from trying new ideas.

While the challenges and difficulties encountered in the exploratory study supported recent findings (Graeske & Sjöberg, 2021; Phoon et al., 2021), specifically those involving the use of VR goggles and internet connectivity, reflecting on them were useful in assessing how VR tours as an online learning supplementary intervention can be replicated successfully in the future. As of this writing, the tours created have been used in two subsequent offerings of the same course, where most of the exploration was presented without VR goggles. However, VR is progressively recognized as a powerful tool for learning and productivity by institutions and governments. Moreover, with some international gatherings like conferences being held in virtual reality, technology companies developing more affordable and portable VR goggles, and groups championing standardization of its use in the educational setting (European Committee for Standardization, 2022), the capability of having the tours viewed using VR goggles by all students will most likely increase should another situation like this arise in the future.

### *5.2. VR tour: Novelty, Presence, and Satisfaction*

Novelty helps with subsequent learning (Fenker et al., 2008; Schomaker et al., 2014). It is also associated with a release of dopamine in the hippocampus (Biel & Bunzeck, 2019), which is related to motivation and pleasure. However, novelty is expected to wane as learners get more familiar with the intervention. The novelty of VR could have contributed to the satisfaction ratings of learners in the first and second lessons. However, the satisfaction ratings and sustained high interest of students in the four succeeding lessons suggested that there was something more than the novelty of VR tours that made them interested not only in the target vocabulary but also in the host country of the language.

An affective aspect of the learning experience made them want to learn more about the country and the culture besides the language. Presence or the *feeling of being in the virtual place* was frequently found to be the reason for students enjoying the activity. The results of the correlation analysis supported this observation by reporting a strong correlation between satisfaction and presence ratings in all five lessons.

### *5.3. Learning Experience more than Outcome*

All fifteen students performed well in the course. Their learning ability as well as the other interventions like the practice exercises, could have mediated actual vocabulary acquisition in this study. Thus, post-test scores were not included. However, the impact of VR tours, as revealed in their interviews, gave us a glimpse of how they would perform in the succeeding courses. Studies have shown that enjoyment in the learning experience is positively related to actual engagement and learning attitudes (Cybinski & Selvanathan, 2005), which lead to better learning outcomes in the long-term.

### *5.4. Immersion, Presence, and Affective Outcomes*

While the relationship between immersion and presence has been well established in the literature (Gorini et al., 2011; Lessiter et al., 2001), the study's findings were insightful because they brought about interesting realizations regarding the objective features of technology (i.e., immersion) and presence being conduits for the affective dimension of the human experience. While immersion was previously found in the literature to be directly influential on presence, the study's findings revealed that the ensuing affective outcomes were only associated with presence implying that the technological features of the device did not solely predict desirable affective outcomes from the multisensorial virtual reality experience. The explanation for this can only be gleaned from further studies. However, based on personal observation, it could be investigated using the lenses of user experience (UX) and ergonomics. Furthermore, the relationship between presence and satisfaction led to the question, "*What makes presence satisfying in a virtual experience?*" While it has been *social presence* that has been found to reduce the adverse effects of physical distance and increase satisfaction among remote learners in past studies (Moore, 2013; Richardson et al., 2017), results from this study suggest that presence experienced in a virtual space regardless of interaction or social connection could increase satisfaction and perceived learning. This can be further investigated by looking into the constructs of interest and sense of place. One could test whether their influence on emotions could be moderated by this spatial form of presence.

## 6. Conclusion and Recommendations

We are ten years away from the deadline of achieving the UN's Sustainable Development Goals, a part of which is achieving quality education for all. In the context of language education, it is our humble belief that achieving quality education cannot be proven by having high scores in exams and tests as proven by the discrepancy of language test scores and actual language learning as exemplified by the study of Nicholson (2015). Instead, we believe that quality education can be achieved by providing enjoyable experiences that establish relevance and authenticity as positive emotions facilitate retention of what students have learned (Dulay & Burt, 1977; Krashen, 1982). In this paper, we presented a new way of learning a foreign language in a remote teaching context through a cognitive innovation based on VR. However, our findings were not without limitations.

The study's lack of generalizability was partly due to the small number of participants. Therefore, an experimental study on testing the effect of immersion on satisfaction levels with larger group sizes is suggested for future researchers. The random assignment of students to groups was limited by the capabilities of their own mobile devices. While this was expected from the study's current context, future researchers may benefit from the fast-paced evolution of technology and recruit participants whose mobile devices would be compatible with VR applications. Furthermore, a face-to-face classroom setting would ensure that everyone would have the chance to use VR goggles.

In conclusion, we hope that the study's contributions, albeit small, would add to the various means of responding to geographical and spatial limitations in education brought about by the pandemic by creating cognitive innovations based on a novel technology like VR. Moreover, the initial findings of the study could support and guide future researchers who are interested in presence, immersion, and other aspects of VR, as well as their impact on learning outcomes. Furthermore, practitioners may benefit from the study as it presented one of the many alternatives and supplements to location-based learning provided by study abroad programs and field trips that would otherwise be inaccessible to more than half of the world's population due to poverty, geopolitical reasons, or disruptions brought about by phenomena like the COVID-19 pandemic. The realizations gleaned from the findings opened up questions regarding immersion's relationship with affective outcomes and suggested lenses for further study. Moreover, the study suggested new hypotheses to be investigated regarding presence and the affective dimension of the human experience.

# Appendices

## Appendix A1

Survey Forms Used for Analysis

One of the main data used in this study was the *after-class survey*. From all the questions, the authors focused mainly on questions number two and three. Question number two was mainly used for quantitative analysis and question number three was mainly used for qualitative analysis.

| **After-class Survey Questions** | |
|---|---|
| **1.名前／ニックネーム**(English: Name/Nickname | **8.ツアー体験した上で、今後今日習った語彙を使う確率はどれくらいだと思いますか？**<br>English: How much do you see yourself using the Filipino words you learned today in the future after the tour?<br>**低い** (Lowest) **高い** (Highest)<br>1 --- 2 --- 3 --- 4 --- 5 --- 6 --- 7 --- 8 --- 9 --- 10 |
| **2.今回の体験を評価してください。**<br>English: How would you rate your experience?<br>**良くなかった** (Lowest)) **良かった** ( Highest)<br>1 --- 2 --- 3 --- 4 --- 5 --- 6 --- 7 --- 8 --- 9 --- 10 | **9.VR ツアーの中で良かったと感じたこと**<br>English: What were the positive feelings you had during the VR tour? |
| **3.問２の評価の理由を述べてください。**<br>English: What's the reason for your rating in number 2? | **10.VR ツアーの中で良くなかったと感じたこと**<br>English: What were the negative feelings you had during the VR tour? |
| **4.ツアー自体はどれくらい面白いと感じましたか？**<br>English: How interested were you in the actual experience?<br>**面白くなかった** (Lowest) **面白かった** (Highest)<br>1 --- 2 --- 3 --- 4 --- 5 --- 6 --- 7 --- 8 --- 9 --- 10 | **11.一つ選んでください** English: Choose One<br>**○VR ツアー中、ただの写真を見ているだけのように感じた**<br>English: During the VR Tour, I felt like I was just looking at a photo.<br>**○ツアーを本当に体験しているかのように感じた**<br>English: I felt like I was in an actual tour. |
| **5.レッスン内容(新しい単語を習うなど)はどれくらい面白いと感じましたか？**<br>English: How much were you interested in the lesson's content (new words)?<br>1 --- 2 --- 3 --- 4 --- 5 --- 6 --- 7 --- 8 --- 9 --- 10 | **12.ただの写真を見ているだけではなく、ツアーを本当に体験しているかのように感じたましたか？**<br>English: How much did you feel that you were in the tour and not just looking at a photo?<br>**そう感じなかった** (Lowest) **そう感じた**(Highest)<br>1 --- 2 --- 3 --- 4 --- 5 --- 6 --- 7 --- 8 --- 9 --- 10 |
| **6.興味を持てた部分は？当てはまるものをすべて選んでください。**<br>English: What were the most interesting parts? | **12.今後の授業で、このようなツアーをもっと体験したいと思いますか？なぜそう思いましたか？**<br>English: Would you like to do more of these tours in future online classes? Why or why not? |
| **7.このツアーを体験した上で、あなた自身が将来今日習ったフィリピノ語の語彙を使うのを想像できますか？どのような場面で英語を使うと思いますか？**<br>English: Do you see yourself using the Filipino words you learned today in the future after the tour? If yes, how? | **13.その他今回の体験に関するコメント、提案、また質問等あればここに書いてください。**<br>English: Please share other comments, suggestions, or questions regarding the whole experience. |

## Appendix A2

Post-semester Survey

The result of the qualitative analysis using question number three of the after-class survey was cross-referenced with the answers of the students in the *post-semester survey*.

### Post-semester Survey

**About Taking Filipino as Major**

*1.Why did you decide to study Filipino?*

*2.What do you want to become in the future (what is your dream job, dream life, etc.)?*

*3.What was your image of the Philippines and of the Filipinos before taking the Filipino class?*

*4.What was your image of the Philippines and of the Filipinos after taking the Filipino class?*

**About Online Class**

*1.How did you like studying online?* **○Not at all ○Very Much**

*2.What are the advantages/merits of studying online?*

*3.What are the disadvantages of studying online?*

*4.How was the length of each class?* **○Very short ○Very long**

*5.If given a choice again, where would you like to have this class, online or in an actual classroom?*

**○Online ○In an actual classroom ○Whichever is fine**

**About the Content of the Class**

*1.In general, how do you like the content of the class?*

**○Not at all ○Very much**

*2.Do you think the themes/topics chosen for this class were appropriate (self-introduction, family, favorite things, birthday, themes of the virtual tours like Rizal Park, Quiapo Church, etc.)?*

**○Not at all ○Very much**

*3. Which theme or topic did you like? Check all that apply.*

**○Self- introduction, Family**
**○Birthday**
**○Favorite Things,**
**○Things inside a room**
**○Describing People**
**○Describing Place,**
**○Rizal Park**
**○Aurora**
**○Quiapo**
**○LRT**
**○Plasa**
**○Museo,**
**○Universities in the Philippines**

**About the Content of the Class**

*4.How much did you understand about the lessons?*

**○Not at all ○Very much**

*5.Which part of the class did you like? Check all that apply.*

**○Grammar explanation**
**○Vocabulary quizzes**
**○Oral recitation (trying to make sentences)**
**○Short class discussion about the theme of the lesson**
**○Reflection each week (Muni-muni)**
**○Virtual Tours,**
**○Active Learning (Video about Learning a Foreign Language, Making Virtual Tours)**

*6.In general, which part of the class is most useful for you? Check all that apply.*

**○Grammar explanation**
**○Vocabulary quizzes**
**○Oral recitation (trying to make sentences)**
**○Short class discussion about the theme of the lesson**
**○Reflection each week (Muni-muni)**
**○Virtual Tours,**
**○Active Learning (Video about Learning a Foreign Language, Making Virtual Tours)**

*7.Which Filipino words or expressions do you think you will remember for a long time?*

*8.What is the most important message or information did you learn in this class?*

*9. Please give any suggestion on how to improve this class.*

## Appendix B

Focus Group Discussion (FGD) Questions

In addition to the answers of the students in the post-semester survey, the result of the qualitative analysis using question number three of the after-class survey was also cross-referenced with the answers of selected students during the two FGDs conducted.

**Focus Group Discussion (FGD) Questions:**

1. なぜ、フィリピン語の勉強を選択したのですか？あなたは、フィリピン語を選択しましたか？
   Why did you decide to study Filipino? Did you choose Filipino or not?
2. フィリピン語の学習にどれぐらいの興味がありますか？１から１０のスケールで、答えてください。
   From a scale of 1 to 10, how interested are you in learning Filipino?
3. フィリピン語の学習に何を期待していましたか？また、その期待は実現しましたか？
   What were you expecting to learn? Did your expectations come true?
4. フィリピン語の授業がオンラインで開かれることを知って、あなたはどのように感じましたか？
   How did you feel when you learned that the Filipino class will be held online?
5. フィリピン語のオンライン授業での最初の１、２日、あなたはどのように感じましたか？むずかしかったですか？簡単でしたか？
   How were your first few days of learning Filipino in an online class? Was it difficult? Was it easy?
6. 授業の中で一番、楽しかった活動は何ですか？また、それは何故ですか？
   What were the most enjoyable activities in the class? Why?
7. フィリピンについて、授業前、どのように考えていましたか？
   How did you think of the Philippines before the class?
8. フィリピンについて、今、どのように考えていますか？
   How do you think of the Philippines now?
9. もし、あなたの見方・考え方が変わったなら、授業の中に見方・考え方を変えるきっかけはありましたか？また、それはどうやって変わりましたか？
   If your view changed, was there something in class that helped you change it? How?
10. VR ツアー以外で面白いアクティビティーはありましたか？
    Is there any activity you had fun except VR tour?
11. バーチャルツアーについてどう思いますか？
    What did you think about the Virtual tours?
12. バーチャルツアーは役に立ちますか？もし、YES ならどのように役に立つと思いますか？
    Were they useful? If yes, how?
13. 一番、好きなツアーはどれでしたか？またそれは何故ですか？
    What was your favorite tour? Why?
14. 一番、面白くなかったツアーはどれですか？また、それは何故ですか？
    What was your least favorite tour? Why?
15. ツアーへの関心は時間と共に変化しましたか？またそれはどのように、何故、変化しましたか？
    Did your interest in the tours change through time? How and why?
16. あなたは、将来、自分がフィリピン語を話している姿をイメージしましたか？
    Did you imagine yourself speaking Filipino in the future?
17. バーチャルツアーはそのイメージを強くしましたか？effect がありましたか？
    Did the virtual tours make your image stronger? Was it effective?
18. バーチャルツアーであなたが困難に感じたことは何ですか？
    What were the problems you encountered with the virtual tours?
19. 来年以降の授業で、バーチャルツアーをおすすめしますか？
    Would you recommend using virtual tours in future classes?
20. その答えについて、何故ですか？?
    Why/Why not?

## Appendix C

Observation Notes

The notes of the authors based on observations during the classes and from the discussion and brainstorming during the weekly researchers' meetings were used to address the challenges encountered during each tour session. The challenges were tackled by adjusting the design and implementation of succeeding lessons.

<table>
<tr><th colspan="2">Observation Notes</th></tr>
<tr><td>TOUR 1: RIZAL<br>1. Started with an orientation on how to use VR Goggles.<br>2. Students had a hard time using the goggles.<br>3. They were excited and eager to do the VR Tours.<br>4. I guess I was not able to clearly explain the groupings because at the end of the class, some students say they didn't know what they should do or should've have done during the first VR Tour.<br>5. Some students were lost.<br>6. Two to three students complained that they did not hear anything.<br>7. Everything took a lot of time – couldn't do the lesson for the day, just the VR Tour. The Tour need more time to explain – maybe next time a separate day for Orientation, Pre-VR Tour Survey, practice, etc.<br>8. Around 2 students request for change of groupings to Group 3 (PowerPoint-based Tour) because their PC couldn't take the tour's memory. One said she is not good with technology so she really didn't know what to do during the whole process.<br>9. No time for after-class survey. We couldn't do it.<br>10. The post-test seemed very easy for them so they all got a perfect score. I think this is because we listened to the audio file several times and the target words were written explicitly in the tour. Maybe the labels should be taken out from the next tour.</td><td>TOUR 2: AURORA<br>1. The new grouping suits the students. The students seem happier.<br>2. Things went smoothly this time.<br>3. We could finish up to after-class survey plus we were able to do the lesson of the day.<br>4. I heard positive feedback from the students today. I think they enjoyed the tour.<br>5. The pre-test and post-test seemed very easy for them.</td></tr>
<tr><td>TOUR 3: QUIAPO CHURCH<br>1. The pre-test and post-test seemed very easy for them. Or they are very good in guessing? Especially for the pre-test or they already know the words from other classes? Some words may have already been introduced in other classes – need to consider this in next experiment.<br>2. A lot of students – G1 and G2 said they liked the photo – very beautiful and it was their first time "to enter" a church.<br>3. One student who was supposed to be in G2 used the VR Tour link for G1. My fault! Need to send link exclusively to each G! students not on the group chat.<br>4. They were shocked to hear/listen to an all-Filipino tour but some students said it was good because they want to listen to more Filipino if possible.</td><td>TOUR 4: LRT<br>1. They also enjoyed this tour because it seemed real.<br>2. They asked why there were only guys on the train. I explained about "women's car".<br>3. I'm not really sure if G1 students are still using the goggles. They have their videos turned off to save internet memory. I couldn't force them to turn it on because of school regulations to respect students' feelings regarding turning on and off videos. What to do about this? For now, just keep on reminding them to use goggles.</td></tr>
<tr><td>TOUR 5: PLASA<br>1. A lot of students enjoyed the photo. They said this and the Quiapo Church are the two most beautiful photos among the tours.<br>2. The Active Learning – VR Tour using PPT by students was a success. Out of the 8 pairs, 3 pairs made their tour using the Filipino language. This was not required, but they said they wanted to try using the Filipino they have learned.<br>3. The tour is going smoothly. We could do lessons and could finish the post-test and after-class survey.</td><td>TOUR 6: MUSEO<br>1. The last tour. Some students who were part of G1 and G2 felt sad about this. Actually, from G3 I didn't hear any feedback all throughout the tour. Well, they said they really didn't want to join because they are not good with pc and that if they joined, they would have had a lot of pc problems.<br>2. They said they learned about the fashion of Filipino people so it was very interesting.<br>3. I discovered that some students didn't answer the survey so I need to give them a follow up on this.</td></tr>
</table>